\documentclass[12pt]{article}
\usepackage{amsmath}
\begin{document} 
\def\a{\alpha}
\def\b{\beta}
\def\c{\varepsilon}
\def\d{\delta}
\def\e{\epsilon}
\def\f{\phi}
\def\g{\gamma}
\def\h{\theta}
\def\k{\kappa}
\def\l{\lambda}
\def\m{\mu}
\def\n{\nu}
\def\p{\psi}
\def\q{\partial}
\def\r{\rho}
\def\s{\sigma}
\def\t{\tau}
\def\u{\upsilon}
\def\v{\varphi}
\def\w{\omega}
\def\x{\xi}
\def\y{\eta}
\def\z{\zeta}
\def\D{{\mit \Delta}}
\def\G{\Gamma}
\def\H{\Theta}
\def\L{\Lambda}
\def\F{\Phi}
\def\O{\Omega}
\def\P{\Psi}
\def\S{\Sigma}

\def\o{\over}
\def\beq{\begin{eqnarray}}
\def\eeq{\end{eqnarray}}
\newcommand{\gsim}{ \mathop{}_{\textstyle \sim}^{\textstyle >} }
\newcommand{\lsim}{ \mathop{}_{\textstyle \sim}^{\textstyle <} }

\renewcommand{\Re}{\mathrm{Re}\,}
\newcommand{\Pf}{\mathrm{Pf}\,}
\newcommand{\Tr}{\mathrm{Tr}\,}
\newcommand{\Kahler}{K\"{a}hler }
\baselineskip 0.7cm

\begin{titlepage}

\begin{flushright}
UT-07-42
\end{flushright}

\vskip 1.35cm
\begin{center}
{\large \bf
Supersymmetric Inflation of Dynamical Origin
}
\vskip 1.2cm
K.~Hamaguchi$^1$, Izawa~K.-I.$^2$, and H.~Nakajima$^1$
\vskip 0.4cm
{\it $^1$Department of Physics, University of Tokyo,\\
     Tokyo 113-0033, Japan}

{\it $^2$Yukawa Institute for Theoretical Physics, Kyoto University,\\
     Kyoto 606-8502, Japan}

\vskip 1.5cm

\abstract{
Dynamical models of inflation are given
with composite inflatons
by means of massive supersymmetric gauge theory.
Nearly flat directions and stable massive ones in the potential
are identified and slow-roll during inflation is examined.
This kind of dynamical inflations may be ubiquitous
in fundamental unified theory with supersymmetry,
which should contain gauge theories for interactions of elementary
particles.}
\end{center}
\end{titlepage}

\setcounter{page}{2}


\section{Introduction}

Cosmological inflation
\cite{Lyt}
is expected to be
a key ingredient to realize our present universe.
Various possible candidates have been considered
as origins of inflaton fields in particle-physics models.
Weakly coupled fields such as moduli fields
in supersymmetric models may be plausible candidates
to yield nearly flat potential for slow-roll inflation.

In contrast, strongly coupled fields might naively
seem inappropriate to achieve flatness of the potential
due to large dynamical corrections to the potential.
However, if dynamical scale as large as fundamental
cutoff scale (or genuinely strong coupling) is to be considered,%
\footnote{In fact, under circumstances such that the dynamical
scale is determined through moduli fixing, it even seems
natural that the dynamical scale is of order the fundamental scale,
while realization of small dynamical scale needs certain tunings
(see Ref.\cite{Izaw}).}
low-energy degrees of freedom may just behave as weakly
coupled fields in effective theory description below the 
cutoff scale, which we assume to be the reduced Planck scale,
for simplicity.

In this paper, we pursue dynamical origin of inflaton fields
by means of composite inflaton models
of massive supersymmetric gauge theory in four spacetime dimensions.%
\footnote{We call such models as dynamical inflation here,
while it can also indicate dynamical origin of inflation scale
(see Ref.\cite{Izaw,Yan}).}
Although the supergravity corrections are important
along with those originated from the strong dynamics,
we ignore them in the present analysis as the first step,
anticipating cancellation among them.
We consider a possibility that such an inflaton is identified
with the primordial one, though primary inflations
\cite{Lyt,Iza}
proceeding the primordial one might be the appropriate stage
for dynamical inflation. Anyway, in both cases, the dynamical inflation
may be relevant for vacuum selection processes in the very early universe.


\section{Preliminary notes}

Let us recapitulate an effective theory description
of supersymmetric $Sp(N)$ gauge theory
\cite{Int}
with $2N+4$ chiral superfields $Q_i$ of the fundamental representation,
where we adopt the notation $Sp(1) = SU(2)$ and $i=1,\cdots,2N+4$
denotes a flavor index with the gauge index omitted.

The effective superpotential
which may describe the dynamics of the gauge interaction
is given by
\begin{eqnarray}
 W_\mathrm{eff} = \frac{1}{\L^{N-1}} \,\Pf M
\label{dynamical_potential}
\end{eqnarray}
in terms of gauge-invariant low-energy degrees of freedom
\begin{eqnarray}
 M_{ij} = -M_{ji} \sim \frac{1}{\L} Q_iQ_j,
\end{eqnarray}
where $\L$ denotes a dynamical scale of the gauge interaction
and $(\Pf M)^2=\det M$.

As for the effective K{\" a}hler potential,
we adopt the minimal one with the higher-order corrections
comparable to the supergravity ones ignored in this paper.


\section{Dynamical inflation}

For concreteness, we first present 
an inflationary model based on
the effective theory description
of supersymmetric $Sp(2)$ gauge theory
with 8 chiral superfields $Q_i$ of the fundamental representation,
where $i=1,\cdots,8$.

\subsection{The model}

Introduction of mass terms for the superfields $Q_i$
yields the effective superpotential for $M_{ij}$ as
\begin{eqnarray}
W_\mathrm{eff} = \frac{1}{\L} \,\Pf M - m_{ij} \L M_{ij},
\end{eqnarray}
where $m_{ij}$ denote mass parameters.
Let us adopt the Planck unit where the reduced Planck scale is unity
and set $\L \sim 1$ as explained in the Introduction.
Without loss of generality, by field redefinitions of the $M_{ij}$,
the mass matrix can be 
block-diagonalized as follows:%
\footnote{Any antisymmetric matrix $A$ can be block-diagonalized
as $A = U\hat{A}U^T$ where $U$ is a unitary matrix and $\hat{A}$
is block-diagonal
\cite{Zum}.}
\begin{align}
         \begin{bmatrix}
           0 &  m_{12} & & & & & & \\
           -m_{12} & 0 & & & & & & \\
           & & 0 &  m_{34} & & & & \\
           & & -m_{34} & 0 & & & & \\
           & & & & 0 &  m_{56} & & \\
           & & & & -m_{56} & 0 & & \\
           & & & & & & 0 &  m_{78} \\
           & & & & & & -m_{78} & 0 
         \end{bmatrix},
\end{align}
where $m_{12}$, $m_{34}$, $m_{56}$, and $m_{78}$
are real and non-negative.

In the following, we consider the case with the common positive mass,
$m_{12} = m_{34} = m_{56} = m_{78} \equiv m^2/2\L$.
The superpotential is then given by
\begin{eqnarray}
W_\mathrm{eff} = \l \,\Pf M - m^2 (M_{12}+M_{34}+M_{56}+M_{78}),
\end{eqnarray}
where $\l$ is a positive coupling of order one.

In order to examine the form of the potential
from the viewpoint of slow-roll inflation,
we concentrate on the block-diagonal
direction in $M_{ij}$, namely, $M_{12}, M_{34}, M_{56}$, and $M_{78}$,
since there are no unstable directions%
\footnote{Although the masses of such stable directions are not necessarily
larger than the Hubble mass during inflation,
we neglect their effects in the following analyses.}
in all the other components
of $M_{ij}$ around this block-diagonal direction (see the Appendix).

In terms of new variables $\phi_i \equiv M_{2i-1,2i}$
with $i=1,\cdots,4$ and the other components vanishing,
the scalar potential is given by
\begin{eqnarray}
 V = \sum_i \left| \l \frac{\phi_1\phi_2\phi_3\phi_4}{\phi_i}
 - m^2 \right|^2
\end{eqnarray}
under the minimal K{\" a}hler potential.

Along the phases of the fields $\phi_i$, the potential has a minimum
at $\arg(\phi_i)=0$, and hence hereafter
we set $\varphi_i = \sqrt{2} \phi_i$ to be positive so that
\begin{eqnarray}
 V = \sum_i \left( \frac{\l}{\sqrt{8}} \frac{\varphi_1\varphi_2\varphi_3\varphi_4}{ \varphi_i} - m^2  \right)^2.
\end{eqnarray}
The supersymmetric vacuum in the chosen direction is given by
\begin{eqnarray}
 \varphi_i = \varphi_0 \equiv \sqrt{2} \left(\frac{m^2}{\l}\right)^{1/3}.
\end{eqnarray}


\subsection{Inflationary dynamics}

Let us regard $\chi = \varphi_4$ as an inflaton
and adopt slow-roll approximation to investigate
the inflationary dynamics.

For $\chi \gg \varphi_0$,%
\footnote{We simply do not consider
the regime $\chi < \varphi_0$ in this paper.}
the other fields are stabilized at the following values:
\begin{eqnarray}
 \varphi_i = \varphi_0 f\left({\chi / \varphi_0}\right);
 \quad
 f(x) = {1\over\sqrt{x}} + {1\over 4x^2}
 + \mathcal{O}\left({1\over x^{3.5}}\right),
\end{eqnarray}
where $i=1,2,3$.
Substituting these values back into the potential,
the scalar potential for $\chi \gg \varphi_0$ is given by
\begin{eqnarray}
V(\chi) \simeq m^4 \left[ 1 - 2\left({\varphi_0\over \chi}\right)^{3 \over 2}
+\mathcal{O}\left(\left({\varphi_0\over \chi}\right)^3\right)\right].
\end{eqnarray}

The inflationary regime is determined by the slow-roll conditions
$\epsilon(\chi) \le 1,\ |\eta(\chi)|\le 1$,
where
\beq
&&\epsilon(\chi) = \frac{1}{2}
\left( \frac{V'(\chi)}{V(\chi)} \right)^2 \simeq {9\over 2} \varphi_0^3 \chi^{-5},
\\
&&\eta(\chi) = \frac{V''(\chi)}{V(\chi)} \simeq {-{15\over 2}}
\varphi_0^{3 \over 2}\chi^{-{7 \over 2}},
\eeq
which are satisfied for $\chi\ge \chi_f \simeq 1.8\varphi_0^{3/7}$.

The field value $\chi_{N_e}$ corresponding to the $e$-fold number $N_e$ is given by
\beq
N_e\simeq \int^{\chi_{N_e}}_{\chi_f} d\chi {V(\chi)\over V'(\chi)}
\simeq
{2\over 21} \varphi_0^{-\frac{3}{2}} \chi_{N_e}^{7\over 2} - {5\over 7},
\eeq
which leads to
\beq
\chi_{N_e}\simeq \left({21\over 2} N_e\right)^{2\over7} \varphi_0^{3\over 7}.
\eeq
The spectral index of the density fluctuations is thus given by
\beq
n_s &\simeq& 1-6\epsilon(\chi_{N_0})+2\eta(\chi_{N_0})
\nonumber
\\
&\simeq& 1 - {10 \over 7N_0}
\eeq
where $N_0$ is the $e$-fold number corresponding to the present horizon.

If this is the primordial inflation,
its scale is determined by the COBE normalization
\beq
\left|\frac{V(\chi_{N_0})^{3 \o 2}}{V'(\chi_{N_0})}\right|
=
\frac{1}{3} m^{2} \varphi_{0}
\left( \frac{\chi_{N_{0}}}{\varphi_{0}} \right)^{\frac{5}{2}}
\simeq 5.3\times 10^{-4},
\eeq
which implies
\beq
m \simeq 1.9 \times 10^{-3}
\left(\frac{50}{N_0}\right)^{\frac{5}{12}}
\left(\frac{1}{\l}\right)^{\frac{1}{12}}.
\eeq
For $\l \simeq 1$ and $N_0 \simeq 50$, we obtain
$m \simeq 1.9 \times 10^{-3}$,
$\phi_0\simeq 0.021$,
$\chi_f\simeq 0.35$,
and
$\chi_{N_0}\simeq 1.2$.
Although the initial value of the inflaton exceeds the dynamical scale $\Lambda$,
or the Planck scale, this could be reduced by means of supergravity corrections, 
which we not discuss in the present analysis.
The reheating via the inflaton decay may also be caused by higher dimensional operators.


\section{Higher-rank models}

For the general case
of massive supersymmetric $Sp(N)$ gauge theory
with $2N+4$ chiral superfields $Q_i$ of the fundamental representation
($i=1,\cdots,2N+4$ with $N$ of order one),
we start from
\begin{eqnarray}
W_\mathrm{eff} = \frac{1}{\L^{N-1}} \,\Pf M - m_{ij} \L M_{ij}.
\end{eqnarray}
Then, as is the case in the previous section corresponding to $N=2$,
we are led to consider
\begin{eqnarray}
W_\mathrm{eff} = \l \prod_i \phi_i - m^2 \sum_i \phi_i,
\end{eqnarray}
as the effective superpotential
in terms of block-diagonal variables $\phi_i \equiv M_{2i-1,2i}$
for $i=1,\cdots,N+2$.
The scalar potential is given by
\begin{eqnarray}
 V = \sum_i \left| \l \frac{\prod_j \phi_j}{\phi_i}
 - m^2 \right|^2
\end{eqnarray}
under the minimal K{\" a}hler potential.

We again set $\varphi_i = \sqrt{2} \phi_i$ to be positive so that
\begin{eqnarray}
 V = \sum_i \left( \frac{\l}{\sqrt{2^{N+1}}}
 \frac{\prod_j \varphi_j}{ \varphi_i} - m^2  \right)^2.
\end{eqnarray}
The supersymmetric vacuum in the chosen direction is given by
\begin{eqnarray}
 \varphi_i = \varphi_0 \equiv \sqrt{2}
 \left(\frac{m^2}{\l}\right)^{1 \o N+1}.
\end{eqnarray}

Let us regard $\chi = \varphi_{N+2}$ as an inflaton.
For $\chi \gg \varphi_0$, the other fields are stabilized
at the following values:
\begin{eqnarray}
 \varphi_i = \varphi_0 f\left({\chi / \varphi_0}\right);
 \quad
 f(x) = x^{-{1 \o N}} + {1\over N^2}x^{-{N+2 \o N}}
 + \mathcal{O}\left(x^{-{2N+3 \o N}}\right),
\end{eqnarray}
where $i=1,\cdots,N+1$.
Substituting these values back into the potential,
the scalar potential for $\chi$ is given by
\begin{eqnarray}
V(\chi) \simeq m^4 \left[ 1
 - 2\left({\varphi_0\over \chi}\right)^{N+1 \over N}
+\mathcal{O}\left(\left({\varphi_0\over \chi}\right)^{2(N+1) \o N}
\right)\right].
\end{eqnarray}

The inflationary regime is determined by the slow-roll conditions
$\epsilon(\chi) \le 1,\ |\eta(\chi)|\le 1$,
where
\beq
&&\epsilon(\chi) = \frac{1}{2}
\left( \frac{V'(\chi)}{V(\chi)} \right)^2
 \simeq 2\left({N+1\over N}\right)^2{1 \o \v_0^2}
 \left({\varphi_0\o \chi}\right)^{4N+2 \o N},
\\
&&\eta(\chi) = \frac{V''(\chi)}{V(\chi)}
 \simeq -2{(N+1)(2N+1) \o N^2}{1 \o \v_0^2}
 \left({\varphi_0 \o \chi}\right)^{3N+1 \over N},
\eeq
which are satisfied for $\chi\ge \chi_f \simeq \varphi_0^{N+1/3N+1}$.

The field value $\chi_{N_e}$ corresponding to the $e$-fold number $N_e$ is given by
\beq
N_e\simeq \int^{\chi_{N_e}}_{\chi_f} d\chi {V(\chi)\over V'(\chi)}
\simeq
{N^2\over 2(N+1)(3N+1)}\v_{0}^{-{N+1 \o N}}
\chi_{N_e}^{3N+1\over N},
\eeq
which leads to
\beq
\chi_{N_e}\simeq \left({2(N+1)(3N+1)\over N^2} N_e\right)^{N\over 3N+1}
 \varphi_0^{N+1\over 3N+1}.
\eeq
The spectral index of the density fluctuations is given by
\beq
n_s &\simeq& 1-6\epsilon(\chi_{N_0})+2\eta(\chi_{N_0})
\nonumber
\\
&\simeq& 1 - {4N+2 \over (3N+1)N_0}
\eeq
where $N_0$ is again the $e$-fold number
corresponding to the present horizon,
while $N$ is the rank of the gauge group $Sp(N)$.

The COBE normalization requires, for the primordial case,
\beq
\left|\frac{V(\chi_{N_0})^{3 \o 2}}{V'(\chi_{N_0})}\right|
=
\frac{1}{2} \frac{N}{N+1} m^{2} \varphi_{0}
\left( \frac{\chi_{N_{0}}}{\varphi_{0}} \right)^{\frac{2N+1}{N}}
\simeq 5.3\times 10^{-4},
\eeq
which implies
\begin{eqnarray}
m&\simeq&
2^{\frac{N+1}{12N}}
\left(
\frac{2(N+1)}{N}
\times 5.3\times 10^{-4}
\right)^{\frac{3N+1}{6N}}
\left(\frac{1}{\lambda}\right)^{\frac{1}{6N}}
\left(\frac{2(N+1)(3N+1)}{N^2}N_0\right)^{\frac{-2N-1}{6N}}\,.
\end{eqnarray}
For $\l \simeq 1$ and $N_0 \simeq 50$, we obtain, for $N=3$, 4, and 5, 
\begin{eqnarray}
m \simeq 2.6 \times 10^{-3}\,,
&
\phi_0\simeq 0.073\,,
&
\chi_f\sim 0.4\,,
\quad
\chi_{N_0}\simeq 2.2
\quad
(N=3)\,,
\\
m \simeq 3.1 \times 10^{-3}\,,
&
\phi_0\simeq 0.14\,,
&
\chi_f\sim 0.5\,,
\quad
\chi_{N_0}\simeq 3.0
\quad
(N=4)\,,
\\
m \simeq 3.5 \times 10^{-3}\,,
&
\phi_0\simeq 0.21\,,
&
\chi_f\sim 0.6\,,
\quad
\chi_{N_0}\simeq 3.6
\quad
(N=5)\,.
\end{eqnarray}


\section{Conclusion}

We have investigated simple%
\footnote{That is, without recourse to certain tunings
such as those in
Ref.\cite{Shi}.}
massive supersymmetric gauge theories
which dynamically provide flat potentials for cosmological inflation.

The $Sp(N)$ gauge group is utilized with $2N+4$ chiral superfields
of the fundamental representation
to result in inflation of the hybrid type with composite inflatons.
More general gauge theories might be of interest along similar
lines, and also inflation of the new type might be realized
around the origin of the field space in certain gauge theories.

The gauge theory is assumed to have genuinely strong coupling
with its dynamical scale around the fundamental cutoff scale
in the supergravity theory where it is immersed.%
\footnote{Such strong coupling may come into play in
circumstances implied, for instance, by heterotic M-theory
with a sizable bulk
\cite{Wit}.}
The expected ubiquity of (massive) gauge theories
in fundamental unified theory with gauge interactions
of elementary particles
may indicate reality of the dynamical inflation,
at least, as primary inflations.

In this paper, we have not analyzed supergravity corrections as well as
dynamical ones in the K{\" a}hler potential,
which should be examined in a future work.


\section*{Acknowledgements}

This work is supported by 
the Grant-in-Aid for Yukawa International Program
for Quark-Hadron Sciences and
the Grant-in-Aid for the 21st Century COE
``Center for Diversity and Universality in Physics''
from the Ministry of Education, Culture, Sports, Science and
Technology (MEXT) of Japan. The work by KH was supported by JSPS (18840012).


\section*{Appendix}

In this Appendix, we show that at the points
along the block-diagonal direction
$M_{12}$, $M_{34}$, $\cdots$, $M_{2N+3,2N+4}$,
there are no unstable directions in all the other components of $M_{ij}$
for the massive $Sp(N)$ gauge theory with the common mass.%
\footnote{$N$ is the rank of the gauge group with $N=2$ in section 2.}

For a superpotential
\beq
W = \l \,\Pf M - m_{ij} M_{ij}
\eeq
with the minimal K{\" a}hler potential,
the scalar potential of the fields $M_{ij}$ is given by
\beq
&&V = \sum_{i<j}\left|\frac{\partial W}{\partial M_{ij}}\right|^2
= \frac{1}{2}\sum_{i,j}\left|
\frac{1}{2}\l \,\Pf M (M^{-1})_{ji}-m_{ij} \right|^2
\nonumber\\
&=& \frac{1}{8}\l^2|\det M| \Tr(M^{-1} {M^{-1}}^{\dagger})
+\frac{1}{2}\Tr(m m^\dagger)
- \frac{1}{4}\l\left[\Pf M \Tr(M^{-1}m^*) + \mathrm{h.c.}\right],
\label{eq:V}
\eeq
where we have used $2 \,\Pf M \, \d (\Pf M) = \d (\det M)
= \det M \, \Tr (M^{-1} \delta M)$.
Note that the parameters $m_{ij}$ here stand for
$m_{ij}\L$ in the main text and $m$ denotes the corresponding matrix.

Now let us decompose the fields $M_{ij}$ into the block-diagonal direction and the other directions:
\beq
M = U \hat{M} U^T,
\eeq
where
\beq
\hat{M} =
\left(
\begin{array}{cc|cc|ccc}
& \phi_1 &&&&&
\\
-\phi_1 &&&&&&
\\
\hline
&&& \phi_2 &&&
\\
&& -\phi_2 &&&&
\\
\hline
&&&& \cdot &&
\\
&&&&& \cdot &
\\
&&&&&& \cdot
\end{array}
\right),
\eeq
and $U$ denotes a unitary matrix, which we can parametrize as
\beq
 U = \exp\left(i\theta_a X_a\right)
\eeq
for the present purposes.
Here $\phi_i$ are positive variables and $X_a$ are hermitian matrices
with real variables $\theta_a$.

We would like to show that the potential has no unstable direction
at $\theta=0$, or the point with all the $\theta_a$ vanishing,
when the block-diagonal direction $\phi_i$ develops
nonvanishing values.
In the scalar potential Eq.(\ref{eq:V}),
only the last term depends on the $\theta_a$:
\beq
V(\theta) = -\frac{1}{4}\l \,\Pf \!\left(U \hat{M} U^T \right)
\Tr(U^* \hat{M}^{-1} U^\dagger m^*) + \mathrm{h.c.}
\eeq

Hereafter, we utilize the form of the common mass
as assumed in the main text, that is,
\beq
m_{ij} = m_0 \O_{ij}; \quad
\O =
\left(
\begin{array}{cc|cc|ccc}
& 1 &&&&&
\\
-1 &&&&&&
\\
\hline
&&& 1 &&&
\\
&& -1 &&&&
\\
\hline
&&&& \cdot &&
\\
&&&&& \cdot &
\\
&&&&&& \cdot
\end{array}
\right)\;,\quad m_0>0.
\eeq

Firstly, the dependence on the overall phase direction
$U=e^{i\theta_0}$ is given by
\beq
V(\theta_0) &=& -\frac{1}{4}\l \,\Pf \hat{M}
e^{2(N+2)i\theta_0} m_0 \Tr(\hat{M}^{-1}\O)e^{-2i\theta_0}
+ \mathrm{h.c.}
\\
&=&
-\frac{1}{2}\l \,\Pf \hat{M}
m_0 \Tr(\hat{M}^{-1}\O)
\cos(2(N+1)\theta_0).
\eeq
Note that $\Pf \hat{M} > 0$ and $\Tr(\hat{M}^{-1}\O)>0$,
and hence $\theta_0=0$ is a minimum.
Thus we take $\det U = 1$ henceforth, which results in
\beq
V(\theta) = -\frac{1}{4}\l \,\Pf \hat{M} m_0
\Tr\left(U^* \hat{M}^{-1} U^\dagger \O \right) + \mathrm{h.c.},
\eeq
since $\Pf(U\hat{M}U^T)=\Pf \hat{M} \det U$.
It is easy to see that $\theta = 0$ is a stationary point:
\beq
\left.\frac{\partial V(\theta)}{\partial \theta_a}\right|_{\theta=0} = 0.
\eeq

In order to examine its stability,
let us decompose the off-diagonal directions
into the following two sets of directions:
$X_a=X_a^{\cal S}+X_a^{\cal A}$, where
\beq
 X_a^{\cal S} = \O X_a^{{\cal S}T} \O^{-1},
 \quad X_a^{\cal A} = -\O X_a^{{\cal A}T} \O^{-1}.
\eeq
It is straightforward to show that
the second derivative with respect to the direction $X_a$
is given by
\beq
\left.\frac{\partial^2 V(\theta)}{\partial \theta_a^2}\right|_{\theta=0}
 = m_0 \l \Tr\left[\hat{M}^{-1}\O \left(X_a^{{\cal S}2}
 + {X_a^{{\cal S}*2}}\right)\right]
 \ge 0,
\eeq
where the equality holds if and only if $X_a^{\cal S}=0$.
Note that the directions with $X_a^{\cal S}=0$
constitute symmetry algebra of the massive gauge theory
with the common mass so that they are completely flat.
Therefore the scalar potential along the block-diagonal direction $\phi_i$
has no instability toward the other directions $\theta_a$.


\end{document}